\documentclass[printer]{aa}

\usepackage[varg]{txfonts}

\newcommand{\subscript}[1]{\textnormal{\tiny{#1}}}
\newcommand{\kms}{km~s$^{-1}$}
\newcommand{\mlr}{M$_\odot$~yr$^{-1}$}
\newcommand{\mjybeam}{mJy~beam$^{-1}$}
\newcommand{\thetapsf}{\ensuremath{\theta_\subscript{PSF}}}
\newcommand{\rleo}{R~Leo}
\newcommand{\nature}{Nature}
\newcommand{\natast}{Nature Astron.}

\renewcommand{\farcs}{\ensuremath{.\!\!^{\prime\prime}}}

\begin{document}

\title{Gas infall and possible circumstellar rotation in R~Leo\thanks{This paper makes use of the following ALMA data: ADS/JAO.ALMA\#2016.1.01202.S. ALMA is a partnership of ESO (representing its member states), NSF (USA) and NINS (Japan), together with NRC (Canada), MOST and ASIAA (Taiwan), and KASI (Republic of Korea), in cooperation with the Republic of Chile. The Joint ALMA Observatory is operated by ESO, AUI/NRAO and NAOJ.}}
\titlerunning{Gas infall and possible circumstellar rotation in R~Leo}
\authorrunning{J. P. Fonfr\'{\i}a et al.}

\author{J. P. Fonfr\'ia\inst{\ref{csic}}
  \and M. Santander-Garc\'ia\inst{\ref{oan}}
  \and J. Cernicharo\inst{\ref{csic}}
  \and L. Velilla-Prieto\inst{\ref{oso}}
  \and M. Ag\'undez\inst{\ref{csic}}
  \and N. Marcelino\inst{\ref{csic}}
  \and G. Quintana-Lacaci\inst{\ref{csic}}}

\institute{Molecular Astrophysics Group, Instituto de F\'isica Fundamental, CSIC, C/ Serrano, 123, 28006, Madrid (Spain);~\email{jpablo.fonfria@csic.es}\label{csic}
  \and
  Observatorio Astron\'omico Nacional, OAN-IGN, Alfonso XII, 3, E-28014, Madrid (Spain)\label{oan}
  \and
Dept. of Space, Earth, and Environment, Astronomy and Plasma Physics Division, Chalmers University of Technology, Onsala Space Observatory, S-439 92 Onsala (Sweden)\label{oso}}

\abstract{
  We present new interferometer molecular observations of R~Leo taken at 1.2~mm with the Atacama Large Millimeter Array with an angular resolution up to $\simeq0\farcs026$.
  These observations permit us to resolve the innermost envelope of this star revealing the existence of a complex structure that involves extended continuum emission and molecular emission showing a non-radial gas velocity distribution.
  This molecular emission displays prominent red-shifted absorptions located right in front to the star typical of material infall and lateral gas motions compatible with the presence of a torus-like structure.
}

\maketitle

\section{Introduction}
\label{sec:introduction}

R~Leo is an O-rich Asymptotic Giant Branch star (AGB) located at $\simeq70-85$~pc from Earth \citep{debeck_2010,ramstedt_2014}.
It has an effective temperature of $\simeq2500-3000$~K and an angular diameter of $\simeq0\farcs025-0\farcs030$ in the K-band \citep{perrin_1999,fedele_2005,wittkowski_2016}.
It pulsates with a period of $\simeq310$~days ejecting processed material with a low mass-loss rate of $\simeq1.0\times10^{-7}$~\mlr{} and a terminal velocity of $6-9$~\kms{} \citep{bujarrabal_1994,cernicharo_1997}.

The continuum at different wavelengths (optical, infrared, mm) together with the thermal and maser molecular emissions coming from this star have been analysed with different observing techniques \citep*[e.g.,][]{castelaz_1997,pardo_1998,ryde_1999,gonzalez-delgado_2003,soria-ruiz_2007}.
Some of these observations suggested the existence of complex structures in the innermost envelope.
\citet{cernicharo_1994} carried out lunar occultation observations of the SiO($v=1,J=2-1$) maser finding a low velocity structure within the 0\farcs5-sized region around the star that could be interpreted as a developing bipolar outflow or a rotating torus.
However, the limited spatial information perpendicular to the occultation direction achieved was insufficient to properly constrain this structure.
Additional asymmetries were found at larger scales by \citet{plez_1994} suggesting that the structure revealed by \citet{cernicharo_1994} could be part of a bigger one covering a significant fraction of the envelope.

In this Letter, we present new interferometer observations of CO and $^{29}$SiO toward the AGB star R~Leo carried out with the Atacama Large Millimeter Array (ALMA) with an angular resolution of up to 0\farcs026, comparable to the size of the star.

\section{Observations}
\label{sec:observations}

\begin{table*}
  \caption{Observations Summary\label{tab:table1}}
  \centering
  \begin{tabular}{c@{\hspace{0.75ex}}c@{\hspace{0.75ex}}c@{\hspace{0.75ex}}c@{\hspace{0.75ex}}c@{\hspace{0.75ex}}c@{\hspace{0.75ex}}c@{\hspace{0.75ex}}c@{\hspace{0.75ex}}c@{\hspace{0.75ex}}c@{\hspace{0.75ex}}c@{\hspace{0.75ex}}c}
    \hline
    Conf./Run & Date        & Phase & \multicolumn{2}{c}{PSF}   & W     & MRS & \multicolumn{2}{c}{Continuum} & \multicolumn{3}{c}{Gaussian Fit}\\
              &             &       & HPBW                      & P.A.  &     &           & Peak              & Integral      & Peak       & D. FWHM                  & D. P.A.\\
              &             &       & (\arcsec$\times$\arcsec)  & (deg) &     & (\arcsec) & (\mjybeam)        & (mJy)         & (\mjybeam) & (\arcsec$\times$\arcsec) & (deg)\\
    \hline
    C40-2     & 2017-Mar-22 & 0.80  & $1.560\times1.080^a$      & 80    & N   & 11.6      & $91.16\pm0.12$    & $105.8\pm0.6$ & ---        & ---                      & ---\\  
    C40-5     & 2017-May-03 & 0.94  & $0.387\times0.359^b$      & 45    & N   & 4.5       & $91.10\pm0.16$    & $111.6\pm1.1$ & ---        & ---                      & ---\\  
    C40-6     & 2016-Oct-01 & 0.25  & $0.186\times0.157^c$      & 32    & N   & 1.6       & $101.0\pm0.3$     & $140.2\pm2.1$ & ---        & ---                      & ---\\  
    C43-9/A   & 2017-Sep-21 & 0.40  & $0.096\times0.035^d$      & 45    & N   & 0.4       & $68.19\pm0.20$    & $103.1\pm1.6$ & ---        & ---                      & ---\\
              &             &       & $0.071\times0.025^e$      & 32    & U   &           & $52.55\pm0.18$    & $103.3\pm1.8$ & ---        & ---                      & ---\\
    C43-9/B   & 2017-Oct-27 & 0.50  & $0.060\times0.046^f$      & 134   & N   & 0.6       & $45.10\pm0.21$    & $90\pm3$      & 37.8       & $0.042\times0.034$       & 173\\  
              &             &       &                           &       &     &           &                   &               & 7.7        & $0.190\times0.021$       & 50 \\  
              &             &       & $0.026\times0.026^g$      & ---   & U   &           & $21.38\pm0.16$    & $85\pm3$      & 16.5       & $0.045\times0.030$       & 155\\  
              &             &       &                           &       &     &           &                   &               & 5.4        & $0.132\times0.022$       & 50\\
    \hline
  \end{tabular}
  \tablefoot{From left to right, the columns contain the ALMA configuration and the run for duplicated observations, the observation date, the optical pulsation phase (pulsation period~$=310$~days; JD$_\subscript{max}=2457896$ following AAVSO\footnote{\texttt{https://www.aavso.org/}}), the size and P.A. of the synthesised PSF, the visibilities weighting (N, ``natural'', means \texttt{robust}~$=1$ and U, ``uniform'', \texttt{robust}~$=0.1$ in GILDAS), the Maximum Recoverable Scale, the Continuum Peak for every configuration, the Continuum Integral above the $5\sigma$-level contours, the Peak Emission of the Gaussian components of the fit, and the FWHM and P.A. after deconvolution.
    If the Gaussian fit is not provided (---) the source can be considered as point-like.
    The errors of the Continuum Peak and the Continuum Integral do not consider the flux calibration uncertainties.
    The statistical errors of the Gaussian fit are always $\lesssim5\%$.
  (\textit{a-g}) The characteristic HPBWs, $\theta_\subscript{PSF}$, are 1\farcs30, 0\farcs37, 0\farcs17, 0\farcs058, 0\farcs042, 0\farcs052, and 0\farcs026, respectively.}
\end{table*}

We observed R~Leo with ALMA during Cycle 4 and 5 under the frame of project 2016.1.01202.S.
Array configurations C40-2, C40-5, C40-6, and C43-9 were used providing baselines from 15~m up to 13.9~km.
These observations give a complete view of the envelope of R~Leo at different scales along $\simeq1.25$ pulsation periods.
The observation details can be found in Table~\ref{tab:table1}.

Four spectral windows covered the frequency ranges $212.7-216.8$ and $227.5-231.5$~GHz with a channel width of 488~kHz.
The flux, bandpass, and pointing were calibrated in the usual way by observing J0854+2006 or J1058+0133.
We estimate a flux error $\simeq5\%$.
The phase calibrators and check sources J1002+1216, J1008+0621, and J0946+1017 were periodically observed.
R~Leo was observed twice with configuration C43-9.
In the first run (run A) the baseline of one spectral window was affected by strong spurious periodic spikes, the weather was worse, and the $uv$ plane coverage was limited.
The data quality highly increased during the second run (run B) and run A was deprecated.
The data calibration was done with the pipeline in CASA~4.7.2 \citep{mcmullin_2007}.

Mapping and data analysis were almost fully performed with GILDAS\footnote{\texttt{https://www.iram.fr/IRAMFR/GILDAS}}.
Imaging restoration was done adopting a robust parameter of 1.0 and 0.1 (natural and uniform weightings, hereafter).
We estimate a positional uncertainty for the highest angular resolution of $\simeq0\farcs005$ from the standard deviation of the emission peak position of the phase calibrator over time \citep{menten_2006,fonfria_2014}.

In this Letter, we have focused on the short scale observations to describe the stellar vicinity.
The analysis of the envelope through the larger scale maps is out of the scope of this work and will be presented elsewhere (Fonfr\'ia et al., \textit{in preparation}).
Due to the stellar pulsation phase incoherence and the long time lag ($\simeq1.1$~yr) the data taken with the C40-6 and C43-9 configurations have not been merged to prevent artefacts.

\section{Results and discussion}
\label{sec:results}

\subsection{The structure of the continuum emission}
\label{sec:continuum}

\begin{figure}[hbt!]
\centering
\includegraphics[width=0.475\textwidth]{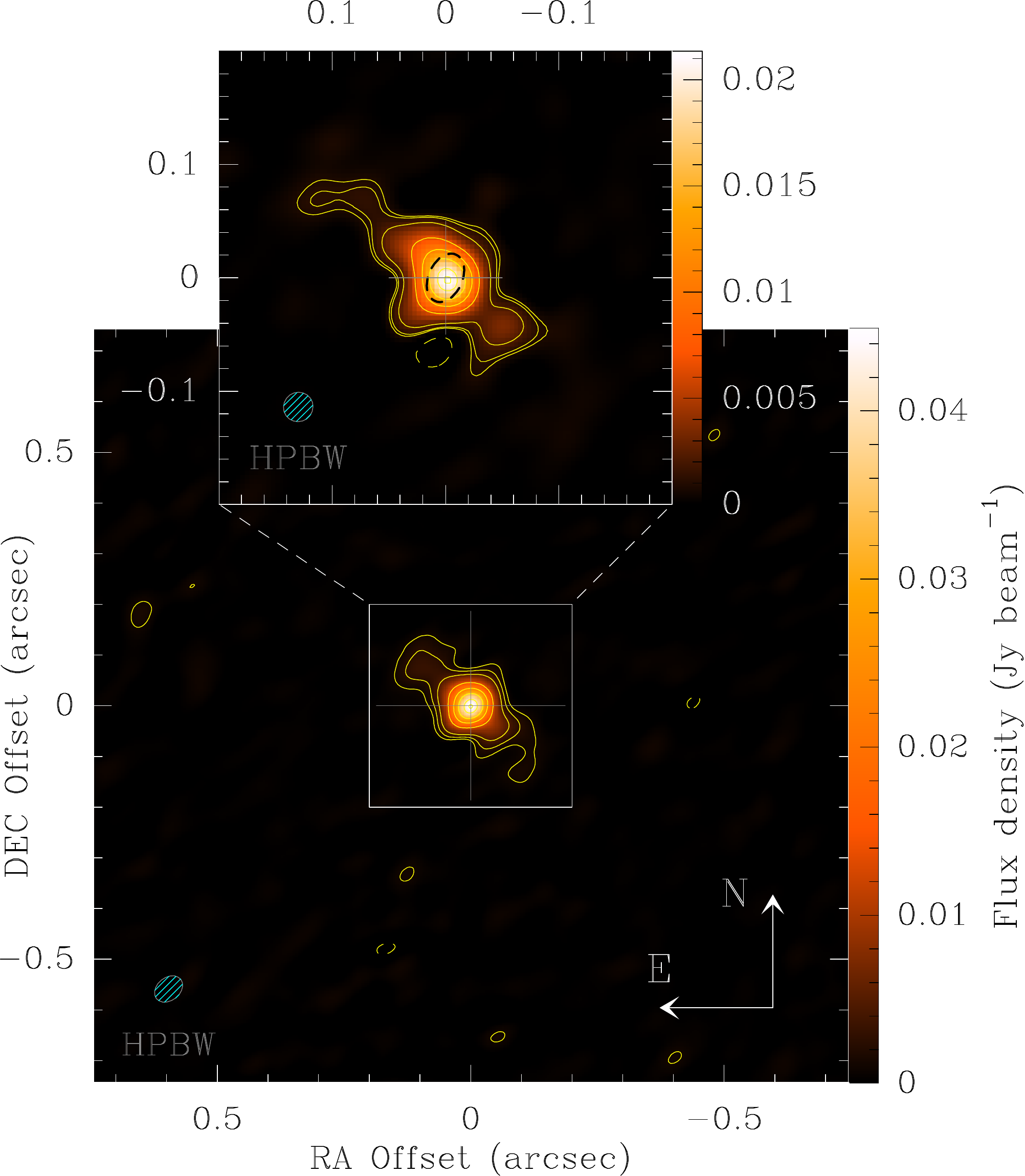}
\caption{Continuum emission observed with configuration C43-9, run B assuming natural weighting (\textit{lower}; $\thetapsf\simeq0\farcs052$) and uniform weighting (\textit{upper}; $\thetapsf\simeq0\farcs026$).
The lowest contours are at $\pm5\sigma$ (2.3\% and 3.75\% of the maximum for the lower and upper insets).
The rest are at 5, 10, 30, 50, 70, and 90\% of the maximum.
The grey crosses indicate where the emission peaks in the uniformly weighted map.
The star is plotted in the upper inset as a dashed ellipse. Its size is the deconvolved size of the compact emission component.}
\label{fig:f1}
\end{figure}

The observations show one continuum emission source in the region of the sky covered by the primary beam (Fig.~\ref{fig:f1}) peaking at $(\textnormal{RA},\textnormal{DEC})=(09^h47^m33.\!\!^s4915\pm0.\!\!^s0004,11^\circ25^\prime 42\farcs899\pm0\farcs005)$.

The highest resolution observations show a compact source surrounded by a faint extended brightness distribution elongated roughly along the Northeast-Southwest direction (Fig.~\ref{fig:f1}).
It is noteworthy that the Maximum Recoverable Scale (MRS; $\simeq 0\farcs6$) is similar to the size of the extended emission in Fig.~\ref{fig:f1}, which is $\simeq0\farcs35$ above the $5\sigma$ level.
Thus, the array could be filtering emission, preventing us to derive reliable brightness temperatures for the star.

The continuum emission can be described by two Gaussian-like components: one compact and one extended (Table~\ref{tab:table1}).
We can estimate the size of the continuum source at 7~mm at the same pulsation phase as during our observations ($\simeq 0.50$) from the results by \citet{matthews_2018} assuming first an uniform disk fit, using linear interpolations, and finally scaling the result to get a Gaussian size.
The result is $\simeq (0\farcs035\pm 0\farcs006)\times(0\farcs025\pm 0\farcs003)$, smaller than the size of our compact source ($0\farcs045\times0\farcs030$).
This only can occur if the star is surrounded by a continuum source that emits more at 1~mm than at 7~mm like a dusty shell \citep{norris_2012}.
All this is compatible with the extended continuum component (size~$\simeq 0\farcs05$) found by \citet{paladini_2017} in the mid-IR around the star.
For simplicity, we hereby consider this compact component as the star.

The extended Gaussian-like component is elongated along the Northeast-Southwest direction.
Its deconvolved size is smaller than the angular resolution provided by the C40-6 configuration ($\thetapsf\simeq0\farcs17$) and it is not evident in the corresponding map.
The ellipticity of this component is $e=1-\theta_\subscript{min}/\theta_\subscript{maj}\simeq0.83$, where $\theta_\subscript{maj}$ and $\theta_\subscript{min}$ are the deconvolved major and minor axes (Table~\ref{tab:table1}).
Since no extended emission has been discovered at longer wavelengths \citep{reid_2007,matthews_2018} the extended component would be produced by dust.

\subsection{Complex molecular line profiles in the vicinity of the star}
\label{sec:complexity}

\begin{figure}[hbt!]
\centering
\includegraphics[width=0.475\textwidth]{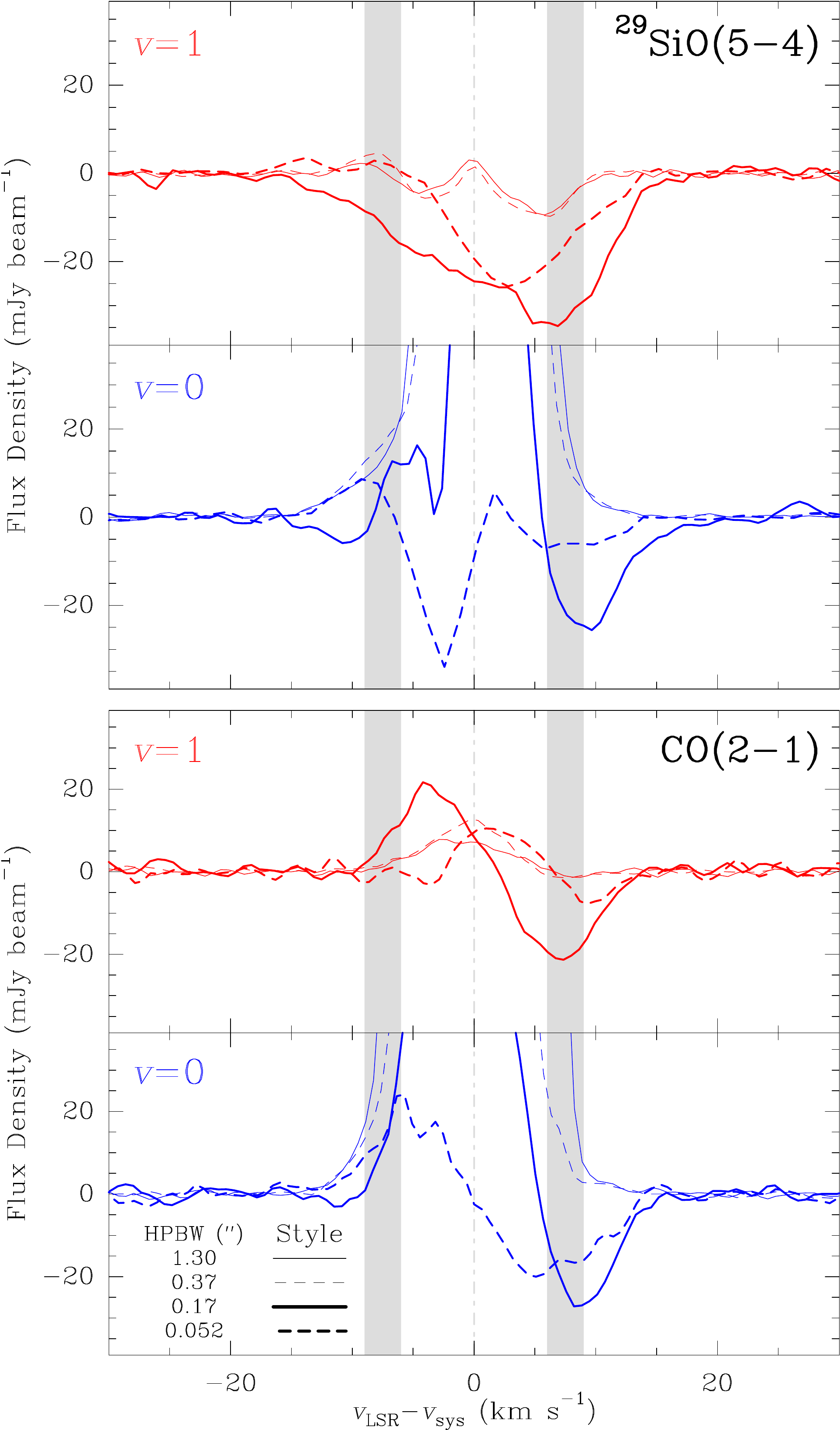}
\caption{Spectra in the central pixel of the CO($2-1$) and $^{29}$SiO($5-4$) lines ($v=0$ in blue and $v=1$ in red) acquired with different ALMA configurations.
  The number of pixels in the PSF is the same for every configuration.
  The grey vertical bands indicate the gas expansion velocity derived from single-dish observations.
  We have assumed a $v_\subscript{sys}=-0.5$~\kms{} \citep{teyssier_2006}.}
\label{fig:f2}
\end{figure}

The CO and $^{29}$SiO lines at the central pixel reveal profiles with emission and absorption components, suggesting the presence of a complex structure (Fig.~\ref{fig:f2}).
The CO($v=0,J=2-1$) and $^{29}$SiO($v=0,J=5-4$) lines observed with an angular resolution $\thetapsf\simeq0\farcs17$ (Table~\ref{tab:table1}) comprise a main emission contribution roughly centred at a systemic velocity of $-0.5$~\kms{} \citep{teyssier_2006} and a prominent red-shifted absorption of $\simeq25\%$ of the continuum peaking at $\simeq8$~\kms.
The $^{29}$SiO line observed with $\thetapsf\simeq0\farcs052$ is dominated by a strong absorption peak at $\simeq-2.5$~\kms{} ($\simeq70\%$ of the continuum) while the CO line shows a clear inverse P-Cygni profile.
Line CO($v=1,J=2-1$) also shows a similar profile with its absorption component peaking at $\simeq10$~\kms.
The $^{29}$SiO($v=1,J=5-4$) line lacks emission but shows a broad absorption peaking at $\simeq3-6$~\kms.
All these absorptions extend up to velocities $\simeq10-15$~\kms, higher than the terminal expansion velocity \citep*[$\simeq6-9$~\kms;][]{bujarrabal_1994,cernicharo_1997,debeck_2010,ramstedt_2014}.
No absorption is observed for all these lines with $\thetapsf\simeq0\farcs37$ and 1\farcs30 except for $^{29}$SiO($v=1,J=5-4$).
This line displays blue- and red-shifted absorptions and a well differentiated, narrow emission component at the systemic velocity.
The large variations seen between the $v=1$ lines, expected to arise very close to the star, observed with $\thetapsf\simeq0\farcs17$ and 0\farcs052 indicate substantial excitation changes throughout a single pulsation period.

All this suggest that
(1) there is gas moving away from us in front of the star producing the red-shifted absorptions,
(2) most of the emission comes from directions that do not enclose the star, and
(3) the upper level involved in the $^{29}$SiO($v=1,J=5-4$) line might be strongly drained, considering that this line forms in shells with temperatures $\gtrsim 1500$~K where the $^{29}$SiO($v=0,J=5-4$) do show a noticeable emission.
This effect could invert the populations of the levels involved in the $^{29}$SiO($v=1,J=6-5$) line, which might show maser emission \citep{cernicharo_1991,gonzalez-alfonso_1997}.

\subsection{Gas infall and hints of rotation}
\label{sec:torus}

\begin{figure*}[hbt!]
\centering
\includegraphics[width=\textwidth]{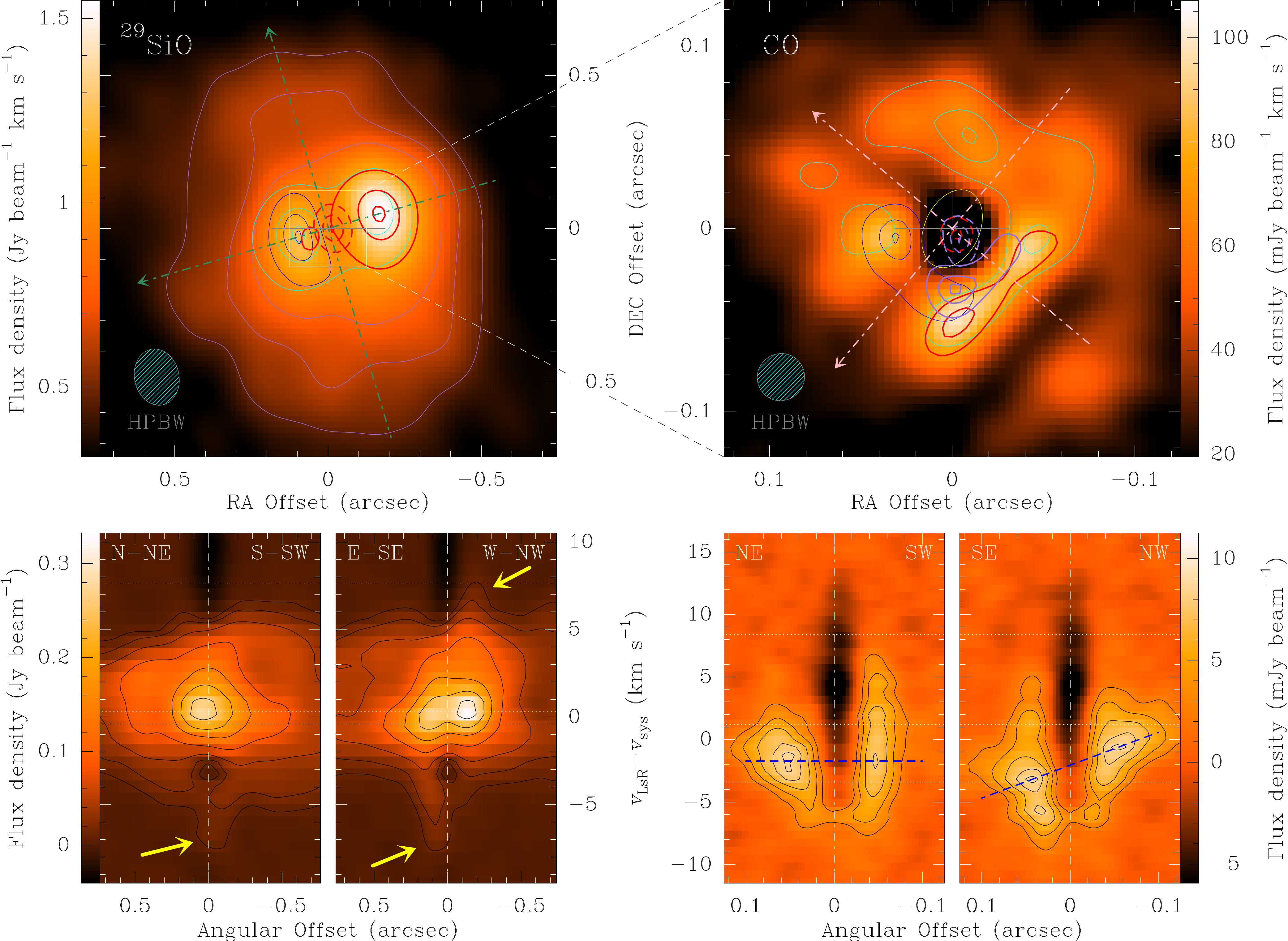}
\caption{$^{29}$SiO and CO emission in the sub-arcsec region.
(\textit{Upper left}) $^{29}$SiO($v=0,J=5-4$) line ($\thetapsf\simeq0\farcs17$).
The purple contours are at levels 30, 40, and 50\% of the integrated emission peak.
The green dashed-dotted arrows describe the main directions of the total emission.
The solid blue, red, and cyan contours are the blue-shifted, red-shifted, and systemic velocity emissions ($[-15,-0.4]$, $[0.3,7.6]$, and $[-0.4,0.3]$~\kms).
The dashed red ones describe the red-shifted absorption ($[7.6,15]$~\kms), which matches spatially up with that of the $v=1$ line.
(\textit{Upper right}) CO($v=0,J=2-1$) line ($\thetapsf\simeq0\farcs026$).
The solid blue and red contours are the blue- and red-shifted emission ($[-15,-3.4]$ and $[1.2,8.4]$~\kms).
The cyan contours are the emission around the systemic velocity ($[-3.4,1.2]$~\kms; $v_\subscript{sys}=-0.5$~\kms).
The dashed red contours are the red-shifted absorption.
The solid and dashed purple contours are the emission and absorption of the CO($v=1,J=2-1$) line ($[-3.3,6.9]$ and $[6.9,15]$~\kms).
The thin yellow ellipse represents the star and the dashed-dotted pink arrows define the main directions of the continuum emission.
In both insets, the contours are at levels 70, 90, and 99\% of the peak emission/absorption.
The grey crosses position the continuum emission peak.
(\textit{Lower left} and \textit{right}) Position-velocity diagrams of the $^{29}$SiO($v=0,J=5-4$) and CO($v=0,J=2-1$) lines (\textit{left} and \textit{right}, respectively) along the corresponding main directions.
The horizontal dotted lines are the limits between the previous velocity ranges.
The yellow arrows (\textit{left}) point out features in the diagrams typical of the emission of rotating gas.
The blue dashed lines (\textit{right}) connect the peak emissions along the Northeast-Southwest and Southeast-Northwest revealing the dependence of the Doppler shift on the direction.
The contours are at 3, 7, 15, 30, 50, 70, and 90\% ($^{29}$SiO; $9.5\sigma\simeq3\%$) and 30, 50, 70, 87, and 97\% (CO; $5\sigma\simeq3.6\%$) of the peak emissions.}
\label{fig:f3}
\end{figure*}

The CO and $^{29}$SiO emission distribution seen in the vicinity of the star are shown in Fig.~\ref{fig:f3}.
The integrated emission of line $^{29}$SiO($v=0,J=5-4$) mapped with $\thetapsf\simeq0\farcs17$ (Fig.~\ref{fig:f3}, upper left panel) shows a bipolar structure laying roughly along the East-West direction (P.A.~$\simeq100^\circ$).
Lower level contours ($\simeq30-50\%$ of the peak emission) are elongated along the perpendicular direction.
The East-West structure can be separated into different contributions in the velocity range $[-15,7.6]$~\kms:
(1) a blue-shifted one centred at $\simeq0\farcs1$ to the East of the star, 
(2) a red-shifted contribution centred at $\simeq0\farcs2$ to the West, and
(3) another contribution at the systemic velocity covering the same region of the envelope as the previous ones.
In the middle of the red- and blue-shifted contributions there is an absorption in the velocity range $[7.6,15]$~\kms, also seen in the $^{29}$SiO($v=1,J=5-4$) line, which matches up with the star within the positional uncertainty.

The division of the CO($v=0,J=2-1$) line into velocity intervals reveals a more complex structure (Fig.~\ref{fig:f3}, upper right panel).
The emission at the systemic and positive velocities of the $v=0$ and 1 lines come from the western hemisphere of the envelope.
The red-shifted absorption of the $v=0$ line matches up with the star but it is slightly shifted toward the Southwest for the $v=1$ line.
The blue-shifted emission of the $v=1$ line is mostly absent.
The blue- and red-shifted emissions of the $v=0$ line and the emission of the $v=1$ line are overlapped to the S.
There is a CO emission deficit at velocities $\gtrsim 3$~\kms{} due Northeast.
The whole picture seems to be roughly symmetric around the Northeast-Southwest direction.
This main direction is compatible with that defined by the CO($v=0,J=2-1$) emission mapped with lower spatial resolutions (Fonfr\'ia et al, \textit{in preparation}) and by the continuum extended emission (Fig.~\ref{fig:f1}).
The position-velocity diagram along the Southeast-Northwest direction (Fig.~\ref{fig:f3}, lower right panel) is similar to those of the rotating disks in expansion found in several post-AGB stars except for the infalling gas \citep{bujarrabal_2016,bujarrabal_2017}.

The disagreement between the symmetry axes in the $^{29}$SiO and CO maps could be consequence of an inhomogeneous chemistry favoured by anisotropic physical and chemical conditions.
Excitation effects cannot be ruled out considering the high number of blended lines shown by SiO and its isotopologues displaying maser emission in this and other O-rich stars \citep*[e.g.,][]{gonzalez-alfonso_1997,pardo_1998}.

Two models can reasonably explain the observations:
\begin{enumerate}
\item Gas rotating around the axis described by the extended continuum emission (Northeast-Southwest direction; P.A.~$\simeq50^\circ$).
  Part of the gas responsible of the blue-shifted emission to the Southeast would leave the main stream during its spinning movement toward the Northwest due to instabilities or interactions with the recently ejected matter, falling onto the star with a Doppler velocity relative to the systemic one up to $\simeq10-15$~\kms{} (Fig.~\ref{fig:f3}, lower insets).
  The rest of the gas would keep revolving around the star at an average velocity of $\simeq6-8$~\kms.
  A similar Doppler shift is expected for gas orbiting a star with a mass of $\simeq 0.5-1.0~M_\odot$ if an orbital inclination of $\simeq 25^\circ-50^\circ$ is assumed.
\item The previous rotating structure would exist as well but there would not be any leakage resulting in gas falling onto the star.
  The red-shifted absorptions would be produced by radial movements of photospheric layers (or convective cells) due to the stellar pulsation \citep{hinkle_1978,hinkle_1979a,hinkle_1979b}.
\end{enumerate}

Contrary to what \citet{vlemmings_2017} claimed to happen in the photosphere of W~Hya, where there is a bright spot suggested to be caused by the collision of infalling matter onto the star, we do not find any footprint of this phenomenon in our continuum maps.
However, the angular resolution of our observations is insufficient to explore the stellar photosphere and a similar bright spot could be highly diluted.
In fact, \citet{tatebe_2008} found evidences of a hot spot in the southern part of R~Leo that might have a counterpart in the mm range.

It can be argued that the Doppler velocities seen in our observations are related to random gas movements in the stellar photosphere or to effects of periodic shocks.
Several works based on VLBI observations of SiO masers have been published for a number of AGB stars such as R~Leo or TX Cam \citep*[e.g.,][]{cotton_2004,soria-ruiz_2007,gonidakis_2013}.
The maser spots are supposed to live enough time to trace the gas movement very close the photosphere, if a good sampling is adopted.
Those observations are constrained to the shell where the maser emission is higher, which covers the region ranging from 0\farcs030 to 0\farcs040.
They thus sample a region of the envelope considerably closer to the star than the one traced by our data.
Common conclusions are therefore difficult to draw.
Contrary to what happens with VLBI observations, the angular resolution of our observations is insufficient to describe random gas movements, which are cancelled out during the averaging performed by the PSF and contribute to broaden line widths.

A set of several clumps ejected with significant axial velocities resulting, for instance, from ballistic trajectories, could also be suggested to comprise the detected structure.
However, our observations show a highly continuous velocity field around the star difficult to explain assuming randomly ejected, discrete clumps with unconnected kinematical properties.
Additionally, there exist evidences of rotating molecular torus in the envelopes of other AGB stars such as L$_2$ Pup or R Dor \citep{kervella_2016,homan_2018}, suggesting that these structures might be more common than initially expected.

The existence of rotating gas in the surroundings of the star needs an angular momentum input, since the rotation of an AGB star is expected to be very low.
This extra angular momentum could be provided by a dipolar magnetic field \citep{matt_2000} or an additional body \citep{vlemmings_2018}.
The first scenario would require a plasma or a gas of charged particles to be coupled with.
No continuum emission typical of a plasma has been detected but dust grains are probably charged in the innermost envelope.
A dipolar magnetic field could force the radially ejected grains to drag the gas inducing circumstellar rotation.
Regarding the second scenario, no companion has been found in the optical and near-IR spectral ranges \citep{gatewood_1992} but this disagreement could be solved if it is a small, cold body such as a low mass star or a Jovian planet instead.
This idea was already invoked by \citet{wiesemeyer_2009b} to explain the polarisation fluctuations detected in the SiO maser emission \citep{wiesemeyer_2009a}, that are also formed very close to the stellar photosphere.
The existence of a planet orbiting an AGB star has been also suggested to be possible by \citet{kervella_2016}, who claim to have discovered what could be one of them in L$_2$ Pup.

\section{Summary and Conclusions}
\label{sec:summary}

We present new interferometer observations of the AGB star R~Leo carried out with ALMA with a maximum angular resolution of $\simeq0\farcs026$ and a MRS~$\simeq0\farcs6$.

The continuum emission can be described as a compact source surrounded by a highly collimated extended emission in the Northeast-Southwest direction.
The CO maps show a 0\farcs1-sized rotating torus-like structure with its axis lying along the Northeast-Southwest direction.
This structure could be explained by (1) either the action of a dipolar magnetic field on the radially expanding wind or (2) the presence of a small, cold companion (maybe a Jovian planet) that is orbiting the central star and injecting angular momentum to the gas.
This seemingly rotating structure is accompanied by evident red-shifted absorptions that could be produced by infalling gas coming from it or by the regular movement of the pulsating photosphere.
At larger scales $^{29}$SiO shows also a torus-like structure with its axis roughly oriented along the North-South direction unseen in the corresponding CO map.
Its origin is unclear but it may be produced by anisotropic physical and chemical conditions, or unexpected excitation effects.

These observations reveal a high complexity structure in the envelope of R~Leo tightly related to the changes undergone by this AGB star.
They will be addressed accurately in the near future through detailed analyses of these and new larger scale data.

\begin{acknowledgements}
We thank the European Research Council (ERC Grant 610256: NANOCOSMOS) and the Spanish MINECO/MICINN for funding support through grant AYA2016-75066-C-1-P and the ASTROMOL Consolider project CSD2009-00038.
MA thanks funding support from the Ram\'on y Cajal program of Spanish MINECO (RyC-2014-16277).
We also thank V. Bujarrabal for his valuable comments about the observations interpretation and the anonymous referee for his/her interesting comments.
\end{acknowledgements}

\end{document}